\documentclass[twoside,slac_one]{revtex4}
\usepackage{a4p}

\usepackage{graphicx}
\usepackage{amsmath} 
\usepackage{bm}
\usepackage{amsxtra}
\usepackage{amssymb}
\usepackage{amsthm}
\usepackage{latexsym}
\usepackage{lscape}
\usepackage{booktabs}
\usepackage{xspace}

\begin{document}

\title{Cosmic Rays from AGN, the Knee Energy Mass Scale and Dark Matter Particles}

%
\author{Yukio Tomozawa}
\affiliation{Michigan Center for Theoretical Physics, Randall Laboratory of Physics,
University of Michigan, Ann Arbor, MI. 48109-1040}
\date{\today }

\begin{abstract}
Following the possibility of a new mass scale at the 3 PeV knee energy of the
cosmic ray energy spectrum, the author suggests that the mass for a dark
matter particle should be 8.1 TeV, using GLMR supersymmetry theory . The
author discusses the possibility of detecting such a signature in various
observational facilities, gamma ray, neutrino and other underground detectors.
An analysis of recent data from HESS yields a gamma ray peak at 7.6 $\pm$ 0.1
TeV, providing observational evidence for a dark matter particle mass
consistent with the theoretical prediction. The author also suggests that the
observed discontinuity in the power law index in galaxy correlation functions
is the knee energy counterpart in cosmology.

\end{abstract}

\maketitle



\section{Introduction}

An extensive search for a dark matter particle (DMP) is under way throughout
the world\cite{dmsearch} by underground detectors using cryogenic or
electronic methods. However, there is no observational hint whatsoever as to
the mass and interactions, etc. of the particle searched for. There is a
slight hope that a forthcoming LHC experiment might give some hint as to the
nature of the particle. That might be wishful thinking, in view of the absence
of any hint from Tevatron experiments in the few TeV\ energy range. The author
will try to construct a scenario for a DMP search, as much as possible based
on the observational data.

\section{High energy cosmic rays from AGN and existence of a new mass scale at
the knee energy}

In a series of articles\cite{cr1}-\cite{cr9} since 1985, the author has
presented a model for the emission of high energy particles from AGN. The
following is a summary of the model.

1) Quantum effects on gravity yield repulsive forces at short
distances\cite{cr1},\cite{cr3}.

2) The collapse of black holes results in explosive bounce back motion with
the emission of high energy particles.

3) Consideration of the Penrose diagram eliminates the horizon problem for
black holes\cite{cr4}. Black holes are not black any more.

4) The knee energy for high energy cosmic rays can be understood as a split
between a radiation-dominated region and a matter-dominated region, not unlike
that in the expansion of the universe. (See page 10 of the lecture
notes\cite{cr1}-\cite{cr3}.)

5) Neutrinos and gamma rays as well as cosmic rays should have the same
spectral index for each AGN. They should show a knee energy phenomenon, a
break in the energy spectral index, similar to that for the cosmic ray energy spectrum.

6) The recent announcement by Hawking rescinding an earlier claim about the
information paradox\cite{hawking} is consistent with this model.

Further discussion of the knee energy in the model yields the existence of a
new mass scale in the knee energy range, in order to have the knee energy
phenomenon in the cosmic ray spectrum\cite{crnew}. The following are
additional features of the model.

7) The proposed new particle with mass in the knee energy range (at 3 PeV) may
not be stable, as in the case of the standard model. If it is a member of a
supersymmetric multiplet and weakly interacting with ordinary particles, then
the stable particle of lowest mass becomes a candidate for a DMP. The only
requirement is that such particles must be present in AGN or black holes so
that the phenomenon of the knee energy is observed when cosmic rays are
emitted from AGN.

8) If the particle is weakly interacting, then it does not obey the GZK
cutoff, since its interaction with photons in cosmic backgroud radiation is
electromagnetic at best. This is a possibe resolution of the GZK
puzzle\cite{gzk},\cite{puzzle}. It will be discussed in a later section.

\section{The knee energy of cosmic ray energy spectrum}

In the traditional theory, cosmic rays below the knee energy are considered to
be of galactic origin, such as those produced by supernova
explosions\cite{hoerandel}, \cite{erlykin}. They are confined inside the
galaxy by the galactic magnetic field for a long time, as far as low energy
components are concerned. The high energy components, however, cannot be
confined inside the galaxy by the galactic magnetic field and therefore they
are considered to\ fall down as a function of energy beyond the knee energy.
This is inferred as the reason for the existence of the knee energy. Then, one
has to introduce, say, extragalactic cosmic rays to supplement the missing
components of the $E^{-3}$ spectrum above the knee energy. A perfect matching
of the intensities of the galactic and extragalactic components of cosmic rays
has to be assumed here. This is the intrinsic difficulty of the traditional
model of cosmic rays, in particular for the explanation of the existence of
the knee energy. This difficulty persists in any model which is constructed
with two independent components for the total cosmic ray spectrum. How does
one find a reason for the matching of intensities for two independent and
unrelated components? The model proposed by the author since 1985 tried to
eliminate such a difficulty. The details are described in the earlier
references\cite{cr1},\cite{cr2},\cite{cr3}. A simple discussion is
recapitulated here.

As described in the previous section, the model started from the realization
that quantum effects on gravity yield a repulsive force at short distances. As
a result, the collapse of a black hole proceeds to an explosion and an
expanding heat bath emits various kinds of particles. This is the reason for
the emission of high energy cosmic rays, gamma rays, neutrinos and dark matter
particles from black holes. The spectrum of an emitted particle $X$ \ with
spin $s$ is calculated by%
\begin{equation}
f_{X}(E)=\frac{2s+1}{2\pi^{2}}\int\frac{E^{2}V_{S}dt}{e^{E/kT-\mu/kT}\pm1},
\end{equation}
where $V_{S}$ is the effective volume around the surface of the heat bath with
temperature $T$\ \ that emits particles. The $\pm$ sign in the denominator is
for fermions/bosons. With the assumption%
\begin{equation}
V_{S}=\frac{4\pi a}{(kT)^{3}}%
\end{equation}
and the expansion rate%
\begin{equation}
t=bR^{\alpha}%
\end{equation}
and%
\begin{equation}
R=\frac{d}{kT},
\end{equation}
where a, b, and d are constants, one can compute the number of particles%
\begin{equation}
f_{X}(E)=\frac{A_{X,\alpha}}{E^{\alpha+1}},
\end{equation}
where%
\begin{equation}
A_{X,\alpha}=\frac{2(2s+1)ab\alpha d^{\alpha}}{\pi}\int_{0}^{\infty}%
\frac{x^{\alpha+2}dx}{e^{x-\mu_{0}}\pm1}%
\end{equation}
and%
\begin{equation}
\mu_{0}=\mu/kT,\text{ \ \ }x=E/kT.
\end{equation}
>From the expansion rate in cosmology, the exponent $\alpha$ can be estimated
as%
\begin{equation}
\alpha=2\text{ \ \ \ \ \ \ \ \ for radiation-dominated regime}%
\end{equation}
and%
\begin{equation}
\alpha=3/2\text{ \ \ \ \ for matter-dominated regime.}%
\end{equation}
This gives the energy spectrum at high energy%
\begin{equation}
f_{X}(E)\approx1/E^{3}%
\end{equation}
and at low energy%
\begin{equation}
f_{X}(E)\approx1/E^{2.5}.
\end{equation}

This is exactly the observed spectrum of cosmic rays. That is the explanation
for the observed cosmic ray energy spectrum and the existence of the knee
energy, proposed in my model in 1985\cite{cr1}..\cite{cr9}. More recently, it
was realized\cite{crnew} that the model requires the existence of a mass scale
at 3 PeV in order to produce the knee energy phenomenon, since without it all
ordinary particles behave as mass-less radiations at temperature 3 PeV. The
existence of a new mass scale is the starting point for the discussion of DMP
in the sections that follow.

Coming back to the discussion of cosmic ray models, gravitaional collapse
yields the emission of high energy particles, cosmic rays, gamma rays,
neutrinos and possibly DMP. The sources can be extragalactic AGN as well as
galactic black holes. Their energy spectra have identical shapes, so simple
addition should yield the final energy spectrum. In order to confront with the
difficulty that galactic cosmic rays fall down above the knee energy, one may
assume that cosmic rays from AGN and black holes dominate the energy spectrum.
The magnitude of galactic cosmic rays produced by supernova explosion is
assumed to be less than 10 \% of the total cosmic ray intensity. Then, the
decrease of the intensities of nuclear cosmic rays will not influence the
total intensity. Of course, this assumption must be carefully examined by the
analysis of future data. Certainly, heavy nuclear components of cosmic rays
satisfiy this assumption. The question whether the proton and helium
components satisfy it must be scrutinized. A complicated structure around the
knee energy observed by the KASCADE group\cite{kascade} may be consistent with
this picture: A small variation in nuclear cosmic rays at this energy can be
the source for complicated structure in the spectrum around the knee energy.

The most important prediction for the author's model is the existence the knee
energy at 3 PeV in the spectra of gamma rays and neutrinos. Such an
observation would clearly show that the knee energy phenomenon has nothing to
do with the galactic magnetic field. One needs to wait a few more years to
test this prediction by observation.

\section{Probability of association between AGN and cosmic ray sources}

In summary, the author's cosmic ray model going back to 1985 has predicted the
Pierre Auger Observatory data\cite{auger}. Moreover it suggests the existence
of a new particle in the PeV mass range, in order to explain the knee energy
phenomenon of the cosmic ray spectrum. There seem to exist some high energy
cosmic ray events that are not associated with AGN among the Pierre Auger
Project data. This is understandable since if AGN-like phenomena are produced
exclusively by dark matter particles (one may call them pseudo-AGN) without a
component of ordinary atomic particles, they wo'nt show up as AGN events. One
needs atomic matter to have an AGN signature. In such cases, cosmic ray events
should be associated with high energy gamma ray emittors. There are gamma ray
emittors that are not associated with known astronomical objects. It is worth
trying to match cosmic ray and gamma ray events with identical unknown
sources. The probability of the association of AGN with high energy cosmic
rays should be of the order of (baryonic matter)/(all matter) = 0.15. Namely,
only 15 \% should have a correlation in a first order approximation. In a
second order approximation, this number might be increased by the emission of
ordinary baryons from exploding black holes. In the long run, baryons emitted
from black holes would increase the probability of converting from a
pseudo-AGN to an ordinary AGN, but it might take a long time. If the
probability of the association of AGN and high energy cosmic rays in the
Pierre Auger Project data were much larger than 15 \%, it would indicate a
high probability of forming ordinary AGN from pseudo-AGN and would further
indicate a probability of forming ordinary galaxies from pseudo-AGN and AGN.

\section{PeV supersymmetry of GLMR and the DMP mass}

In order to have a mass scale of 3 PeV and DMP of relatively low mass, one has
to have a supersymmetry model with a large mass ratio. Such a theory has been
proposed by GLMR (Giudice, Luty, Murayama and Rattazzi). Assuming the absence
of singlets, GLMR derived a large mass ratio\cite{glmr},\cite{pevss}%
\begin{align}
M_{2}  &  =\frac{\alpha}{4\pi\sin^{2}\theta_{W}}m_{3/2}\\
&  =2.7\text{ }10^{-3}m_{3/2},
\end{align}
among other parameter relations, where $\alpha$ and $\theta_{W}$ are the fine
structure constant and the weak interaction angle respectively. Here $M_{2}$
and $m_{3/2}$ stand for a gaugino and a gravitino mass for in the GLMR theory,
respectively. Since this is the largest mass ratio obtained, one may choose
the highest mass scale to be%
\begin{equation}
m_{3/2}=3\text{ }PeV,
\end{equation}
then one gets%
\begin{equation}
M_{2}=8.1\text{ }TeV,
\end{equation}
which is the mass of the lowest mass particle (LMP), i.e., the DMP mass. The
accuracy of the prediction is in the range of 10\symbol{126}20 \% from the
determination of the cosmic ray knee energy. Being weakly interacting, this
particle must be produced by a pair in an accelerator experiment. This makes
it impossible to discover such particles directly in LHC experiments at the
presently planned energy scale. Since, however, this DMP mass will be close to
the maximum energy of LHC experiments in the near future, it might be possible
to see the existence of such a particle as a new physics signature in a LHC
experiment. It is worth noting that if something like the GLMR supersymmetry
is not used, then the DMP mass should be much higher than 8.1 TeV, insofar as
it is related to the cosmic ray knee energy mass scale.

Although knee energy particles are produced at 3 PeV at AGN, they eventually
decay into DMP of 8.1 TeV, and that is what will be observed on Earth. We
consider the direct detection of DMP in cosmic rays, gamma rays and neutrinos
in subsequent sections. Using the name cion (originally sion) for particles in
generic sense\cite{crnew}, one may call the particle at 3 PeV a prime-cion or
urcion, since it is of primary importance. The DMP at 8.1 TeV may be called a
dm-cion for obvious reasons. The name cion comes from the Chinese word for
knee, Xi, pronounced shi. It is also an acronym for cosmic interphase
particle. The cion is introduced to explain the cosmic ray knee energy by a
mass scale that separates the radiation- and matter-dominated phases for black
hole expansion.

The same phenomenon can apply to the expansion of the universe: The different
expansion rates in radiation- and matter-dominated pahses of the universe
should be separated by a temperature of 3 PeV with the introduction of the
prime-cion. This phenomenon may have contributed to the dominance of dark
matter over baryons in the universe. It is worth searching for observational
evidence for different expansion rates in the early universe. In a later
section, the counterpart of the knee energy in cosmology will be sought in the
galaxy correlation functions.

\section{Production of DMP along with cosmic rays and its interaction}

With center of mass energy $m_{3/2}=3PeV$, the interaction becomes maximum at
the lab energy
\begin{align}
E_{\max}  &  =\frac{m_{3/2}^{2}}{2M}=0.32\text{ }10^{21}eV=0.32\text{ }ZeV\\
\text{ }for\text{ \ }M  &  =14\text{ }GeV,
\end{align}
where the cross section is of the order of electromagnetic interactions. A
mass scale of 14 GeV has been chosen from a dominant mass in the atmosphere,
that of nitrogen. This energy is close to the GZK (Greisen, Zatsepin and
Kumin\cite{gzk}) cutoff energy. The DMP cross section decreases linearly with
energy, while the intensity increases with decreasing energy, ($E^{-3}$ above
the knee energy and $E^{-2.5}$ below the knee energy.) Therefore, the product
of the cross section and the intensity function favors the low energy end for
detection. In other words, DMP in cosmic rays accumulate on the earth and the
lowest end of the spectrum can be observed most easily. This tells what is the
best way to detect DMP in cosmic rays.

The cross section for DMP interaction can be parametrized as%
\begin{equation}
\sigma=\sigma_{EM}(E/E_{\max}),
\end{equation}
where%
\begin{equation}
\sigma_{EM}=10^{-26}cm^{2}=10^{-30}m^{2}.
\end{equation}
Assuming that the energy distribution of DMP from AGN is the same as that of
observed cosmic rays, one can parametrize it as%
\begin{equation}
F=\frac{\sqrt{2}m_{3/2}^{3}}{E^{5/2}(E+m_{3/2})^{1/2}}F_{KN},
\end{equation}
where%
\begin{equation}
F_{KN}=3.0\text{ }10^{-14}(m^{2}sr\text{ }s\text{ }GeV)^{-1}%
\end{equation}
stands for the flux at the knee energy, $m_{3/2}=3PeV$. The resultant flux for
target mass $M$ in one km w.e. (water equivalent),%
\begin{equation}
\rho=(\frac{1}{1.67\text{ }10^{-30}}\ )10^{3}\frac{1GeV}{M}\text{ }%
m^{-2}=0.59\text{ }10^{33}\frac{1GeV}{M}\text{ }m^{-2}%
\end{equation}
is given by%
\begin{align}
I  &  =%
{\displaystyle\int\limits_{M_{2}}^{\infty}}
F\sigma\rho dE=\frac{F_{KN}\sqrt{2}m_{3/2}^{3}\sigma_{EM}\rho}{E_{\max}}%
{\displaystyle\int\limits_{M_{2}}^{\infty}}
\frac{dE}{E^{3/2}(E+m_{3/2})^{1/2}}\\
&  =\frac{2\sqrt{2}m_{3/2}^{2}\sigma_{EM}\rho}{E_{\max}}\sqrt{\frac
{M_{2}+m_{3/2}}{M_{2}}}\\
&  =3.0\text{ }10^{-14}\ast4\sqrt{2}\text{ }10^{-30}\ast0.59\text{ }%
10^{33}\sqrt{m_{3/2}/M_{2}}\\
&  =1.92\text{ }10^{-9}m^{-2}sr^{-1}s^{-1}. \label{eq01}%
\end{align}
The final result is independent of the target mass, M, and is proportional to
$\sqrt{m_{3/2}/M_{2}}$. The value of this quantity has been chosen to be%
\begin{equation}
\sqrt{m_{3/2}/M_{2}}=1/\sqrt{2.7\text{ }10^{-3}}=19.2
\end{equation}
in the above estimate. The choice of a smaller value for $M_{2}$ results in an
increased value for Eq. (\ref{eq01}).

\section{Underground muons, a bump in the dip}

The vertical muon intensity for underground muons as a function of depth
(measured in the units of km.w.e.) is well investigated\cite{crdata}. Cosmic
ray muons impinging on the surface of the Earth ground level start to decrease
by scattering and decay, and then become a flat distribution at a depth of 15
km.w.e. The vertical muon intensity is given as the saturated value of the
vertical intensity with%
\begin{align}
F_{\mu}  &  =4.0\text{ }10^{-9}\text{ }m^{-2}sr^{-1}s^{-1}\text{
\ \ \ \ \ \ \ }\label{eq02}\\
&  =2.0\text{ }10^{-9}\text{ }m^{-2}sr^{-1}s^{-1}\text{ \ \ \ \ \ \ \ .}
\label{eq03}%
\end{align}
in the horizontzl and vertically upward directions respectively. The
interpretation of this data is that the flat distribution is due to muon
production by atmospheric neutrinos, which are the product of cosmic rays
hitting the atmosphere. The difference between horizontal and vertical in Eqs.
(\ref{eq02}) and (\ref{eq03}) is considered to be due to neutrino oscillation.
Depending on the direction and energy of the injected atmospheric neutrinos
and the location of the interaction, the intensity of neutrinos and
subsequently produced muons varies by neutrino oscillation.

The vertical muon intensity varies as%
\begin{equation}
F_{\mu}=Ae^{-bX}%
\end{equation}
for small depth $X$ measured in units of km.w.e.. From Figure 20.5 of Ref.
\cite{crdata}, one can read%
\begin{equation}
\frac{F_{\mu}(X=10)}{F_{\mu}(X=1)}=e^{-9b}=10^{-6},
\end{equation}
and then%
\begin{equation}
b=\frac{6}{9}\ln10=1.54\text{ \ \ }(km.w.e.)^{-1}.
\end{equation}
The dark matter contribution for muon production is computed by%
\begin{equation}
\frac{dF_{\mu}(DMP)}{dX}=a-bF_{\mu}(DMP), \label{eq04}%
\end{equation}
where%
\begin{equation}
a=\epsilon_{\mu}I.
\end{equation}
Here $I$ is computed in Eq. (\ref{eq01}) and $\epsilon_{\mu}$ is the muon
multiplicity. Since the process is dominated by the low energy end of the dark
matter interaction, one can assume that%
\begin{equation}
\epsilon_{\mu}=1.
\end{equation}
The solution of Eq.(\ref{eq04}) is given by%
\begin{equation}
F_{\mu}(DMP)=\frac{a}{b}(1-e^{-bX}),
\end{equation}
where the asymptotic value of $F_{\mu}(DMP)$\ is%
\begin{equation}
F_{\mu}(DMP)=a/b=I/b=1.25\text{\ }10^{-9}m^{-2}sr^{-1}s^{-1}.
\end{equation}
Using Eq.(\ref{eq02}), one gets a separation of neutrino and DMP contributions
for the asymptotic value,%
\begin{equation}
\frac{F_{\mu}(neutrino)}{F_{\mu}(DMP)}=\frac{4.0-1.25}{1.25}=\frac{2.75}%
{1.25}=2.2.
\end{equation}
In other words, we have concluded that the contribution of DMP for vertical
muons is 45 \% of that for neutrinos.

Based on this estimate, one can propose a method of detecting the DMP
contribution for vertical muons. First choose the direction for which the
damping effect due to neutrino oscillation is maximal. This is the direction
which creates maximum suppression of the peak in the muon momentum
distribution. In other words, one should observe a deep dip in the muon
momentum distribution. Since there is no oscillation phenomenon for DMP, the
contribution of DMP to vertical muon production is determined as a bump in the
dip. This is the key for the discovery of a DMP contribution in the muon
distribution. In any other direction, there is a tilted excess over a dip.

Let me mention the status of the AMANDA observation\cite{amanda}. The data
relevant for the DMP mass covers a range below 6 TeV at the present time,
short of 8 TeV. The absence of significant DMP evidence in AMANDA is
inevitable. Relevant data on a bump in the dip in AMANDA is given in Fig. 2,
where the zenith angle distribution of the upward muon is shown. The number of
events in one direction is of the order of 10. This is not enough events to
see the phenomenon of a bump in the dip. One may need one order of magnitude
more events. The same comments can be applied to other underground detectors,
such as super-Kamiokande, Sudan etc. One may need a larger detector, such as
DUSEL (Deep Underground Science and Engineering Laboratory), planned in
Homestake, SD.

\section{A sharp peak in gamma ray spectrum}

The collision of a DMP and its antiparticle produces 2 gammas,
\begin{equation}
DMP+AntiDMP\rightarrow\gamma+\gamma.
\end{equation}
peaked at the DMP mass. In our model this sharp peak should occur at 8.1 TeV.
This is in an accessible range for present gamma ray observatories such as
HESS and VERITAS. Obviously, this is an event similar to the observation of
511 keV gamma rays as evidence for positrons. The author will present a
discussion about where to look for such events.

\subsection{PKS 2005-489}

HESS observed gamma ray events above 100 GeV from BL Lac object PKS 2005-489
(z = 0.071) in 2004-2007. A summary of their observation reads\cite{hess}

1) Below 2 TeV, they obtained the spectrum of a power law fit with a photon
index of 3.2.

2) Above 2 TeV (upto 10 TeV), they observed a gamma ray excess relative to the
above power law spectrum.

It would be interesting to see whether this excess would result in a sharp
peak in the multi TeV range upon statistical improvement in the future.

\subsection{Galactic center}

There is a black hole of 2.6 million solar mass at the center of the Milky
Way\cite{gcbh}. Using the model proposed by the author, one can expect the
emission of cosmic rays, gamma rays and neutrinos as well as DMP from such
black holes. This has an advantage for their detection by proximity. While
high energy cosmic rays are diverted by galactic magnetic fields, all of the
other particles mentioned above can be detected directly. In fact, the author
has suggested that the DAMA data\cite{dama} implies that 30 \% of any dark
matter observed there should originate from the galactic center\cite{2comp}.
It would be worthwhile to see whether the remnant of gamma rays from the
galactic center can be observed.

\subsection{Sources of high energy cosmic rays observed by the Pierre Auger
Project}

The gamma rays observed from PKS 2005-489 (z = 0.071) by HESS suggest that TeV
gamma rays can reach the Earth if the sources are relatively nearby.
Therefore, the sources of high energy cosmic rays observed by the Pierre Auger
Project can be attractive for TeV gamma rays and DMP, if the distances are
nearby\cite{auger}. The Pierre Auger Project reported a high concentration of
high energy cosmic ray sources near the location of Centaurus A (l = 309.5158,
b = 19.4173). The distance to Centaurus A is 547 km/s or z = 0.001825. The
abundance of high energy cosmic rays makes this direction an attractive target
for gamma ray observation as well. Why does this direction have copious
sources of high energy cosmic rays? The next subsection will provide a hint.

\subsection{Center of the universe}

The author has determined the location of the center of expansion of the
universe from the observed values of the cmb (cosmic back ground radiation)
dipole and peculiar velocity. The latter is the sum of that of the solar
system towards the Virgo cluster and that of the Virgo cluster towards the
Great Attractor. The observed cmb dipole and total peculiar velocity are very
different both in magnitude and direction, as opposed to the assumption often
made. Based on this observation, the author computed the location of the
center of the universe to be\cite{ucenter}%
\begin{equation}
v=5325.8\pm198\text{ }km/s,\text{ \ }l=313.2\pm0.2^{\circ},\text{ \ }%
b=12.5\pm0.3^{\circ}%
\end{equation}
or%
\begin{equation}
v=5434.5\pm208\text{ }km/s,\text{ \ }l=313.0\pm0.2^{\circ},\text{ \ }%
b=16.4\pm0.3^{\circ}.
\end{equation}
(These are corrected values from a numerical mistake in the values quoted in
the reference\cite{ucenter}. See errata to appear.) The difference between
these estimates originates from two different estimates of the peculiar
velocity for the solar system towards the Virgo cluster by
Sciama\cite{sciama1},\cite{sciama2},\cite{sciama3}. The directions of these
solutions are close to that of Centaurus A The direction of Centaurus A is
nothing but the direction of the center of the universe. It is a special
direction, indeed. Since the model proposed by the author can be applied to
the expansion of the universe, DMP can be emitted when the temperature of the
universe is 3 PeV. This may have something to do with the dominance of dark
matter in the universe over baryons, as was mentioned earlier. Then, remnants
of the expansion accumulated in the neighborhood of the center of the
universe. It would be worthwhile to see whether a sharp peak for gamma rays in
the multi TeV range can be observed in this direction. The distance to the
center of the universe is much closer than that to PKS\ 2005-489, which is
21285 km/s.

\subsection{Statistical sum of all data in the multi-TeV range}

An alternative general suggestion is that one could accumulate all gamma ray
data in the muti-TeV region with an appropriate statistical weight. That could
make it easier to see a sharp peak. After finishing the original version of
this article, I have looked through all the data from HESS, VERITAS, MAGIC and
CANGAROO, and I found one data set which gives a satisfactory answer to this
question. See the following section for the data analysis.

\section{Observed gamma ray peak as evidence for a dark matter particle}

In a recent HESS report\cite{hess8}, high energy gamma rays from 8 unknown
sources have been recorded. The data from each source cover the energy range
of 1 to 40 TeV and have similar statistics, since they have been obtained in a
recent systematic survey. That the sources are unknown may not be a drawback
for a dark matter gamma ray search, since unknown sources may not be ordinary
AGN or other known astronomical objects. If the source is an AGN type object
consisting entirely of dark matter particles (called pseudo-AGN in Section
IV), it may not have the signature of an ordinary AGN, since such a signature
needs ordinary matter to emit atomic photons. The presence of abundant dark
matter favors 2 gamma ray emission from DMP and anti-DMP annihilation. One
does not need to exclude gamma ray emitters such as ordinary AGN etc, since
one expects a DMP environment in such a case as well. The simple sum of gamma
rays from the 8 sources is plotted in Fig. 1. The values at energy 2, 4, 6, 8,
10 and 12 TeV are estimated from the interpolation of those at the neighboring
observed points. The error bars are estimated from the existing data. The sum
clearly shows a peak at 7.6 $\pm$ 0.1 TeV. This is consistent with the
predicted value of 8.1 $\pm$ 0.8 TeV (assuming 10 \% accuracy for the knee
energy determination). See a separate report for more
discussion\cite{evidence}. I did not include any analysis at energy higher
than 12 TeV, since the data points there are in the range of a 10 TeV bin,
while it is a 2 TeV bin below 10 TeV.

If this result were confirmed by further data from HESS and other high energy
gamma ray detectors, that would suggest that LHC experiments will find neither
a dark matter particle nor any supersymmetric new particle with the presently
planned energy scale. One needs 16 TeV to produce a pair of dark matter
particles, each having 8 TeV. A dark matter particle is the lowest mass member
of any supersymmetric theory. The presently attained energy at LHC is 7 TeV,
possibly reaching 14 TeV in the near future. It would be vitally important
if\ a method could be found to increase the output energy of the particle accelerator.

\section{Neutrinos from AGN: Energy spectrum and the knee energy}

As the author has suggested since 1985, neutrinos and gamma rays shoud be
emitted from AGN, for the same reason that high energy cosmic rays are
emitted. The ratio of neutrinos and gamma rays is given by%
\begin{equation}
\frac{A_{\nu,\lambda-1}}{A_{\gamma,\lambda-1}}=\frac{2\ast3\sum_{n=1}^{\infty
}(-1)^{n}\frac{\Gamma(\lambda+2)}{n^{\lambda+2}}}{2\sum_{n=1}^{\infty}%
\frac{\Gamma(\lambda+2)}{n^{\lambda+2}}}=3(1-\frac{1}{2^{\lambda+1}}),
\end{equation}
where $\lambda$ is the gamma ray energy spectrum index. The same formula in
ref. \cite{crnew} is missing a factor of \ 2, which is the formula for a
Majorana neutrino. This value is given as%
\begin{equation}
3(7/8,\text{ }0.912,\text{ }15/16)\text{ \ \ for \ \ }\lambda=(2,\text{
}2.5,\text{ }3).
\end{equation}

The energy spectra for neutrinos and gamma rays emitted from AGN should have
the knee energy phenomenon at 3 PeV, the same energy as that for cosmic rays.
This is because internal structure from the presence of the prime-cion is the
cause of the knee energy. There should be a universal knee energy at 3 PeV for
all particles emitted from AGN. For gamma rays, the energy spectrum may be
modified by interaction with other photons or intergalactic materials, but a
discontinuity in the spectral index should persist.

\section{The AGASA data on GZK cutoff violation}

High energy cosmic rays traversing intergalactic space suffer the GZK
cutoff\cite{gzk} above 100 EeV due to interactions with cosmic background
radiation, if the primary cosmic ray particles are protons or nuclei. The
Pierre Auger Project\cite{auger}, HiRes\cite{hires} and Yakutsk\cite{yakutsk}
found a GZK cutoff, while Akeno-AGASA\cite{agasa} observed the events above
the cutoff (11 events in the past 10 years)\cite{puzzle}. Since the number of
events that violate the GZK cutoff has been steadily increasing in the past 10
years, the discrepancies among the results for different detectors must be
explained by experimentalists. Since the result of the Akeno-AGASA experiment
is smooth near the cutoff energy, we have to accept their result and wait for
a future explanation of the differences among the detectors. The author will
assume that the Akeno-AGASA result is correct and consider its implication,
until otherwise noted.

An important difference between the AGASA instrument and most of the other
cosmic ray detectors is the capability of gamma ray shower
observation\cite{agasa}. Because of the small scale of the detector and direct
measurement of the shower, AGASA was able to detect a lot of gamma ray
showers, and GZK-cutoff-violating showers are on the borderline between gamma
ray showers and ordinary nuclear showers. Therefore, it is not surprising that
only the AGASA detector has observed GZK-cutoff-violating events. One has to
have the capability of observing gamma ray showers in order to catch
GZK-cutoff-violating events. In this respect, a new project of the Utah group
(TAP, Telescope Array Project\cite{tap}) that aims to detect gamma ray showers
is the most promising approach.

A possible explanation for the AGASA data on GZK cutoff violation would be a
shower caused by a DMP. A DMP is not constrained by the GZK cutoff, since it
interacts weakly with cosmic background radiation. Then the question is how
such a particle can be accelerated to an energy as high as 100 EeV. The model
described in this article and since 1985 answers precisely this question. As
is described in the earlier sections, the production of DMP, as well as high
energy cosmic rays, from AGN is a natural conclusion of the model.

\section{The remnant of SN87A}

Many remnants of supernovae II are neutron stars or pulsars. The remnant of
the supernova SN87A is not a pulsar. Although there is no evidence reported
yet, it is most likely a black hole, if not a neutron star with a slow
rotation. If it is a black hole, it will never be a neutron star in the future
according to the old theory of black holes, since the mass of a black hole can
only increase, never decrease. However, with the model proposed by the author
since 1985, the mass of a black hole can decrease by the emission of cosmic
rays, gamma rays, neutrinos and DMP. Black holes are not black any more. Then,
a black hole, a remnant of SN87A, can become a neutron star, if some matter
such as a stray astronomical object collides with it, resulting in an
explosion. It could lose enough mass to become a neutron star. At least that
is a possible scenario.

\section{Implication for cosmology: galaxy correlation functions}

The scenario for black hole explosion by quantum-field-theory originated
repulsive forces can be applied to the expansion of the universe. In
particular, if the expansion of the universe is preceded by its collapse, then
the both scenarios for black holes and the universe are identical. Therefore,
the split between radiation-dominated and matter-dominated eras should occur
at a temperature of 3 PeV for the universe, similar to that for AGN. The
presence of prime-cion and dm-cion must be the cause of the predominant
presence of dark matter in the universe. Then, can one find evidence for a
discontinuity in the index of the expansion rate of the universe, just as the
knee energy for cosmic rays is evidence for a difference in the expansion
rates for AGN? Where is the history of the expansion of the universe
imprinted? The most likely place would be in the galaxy correlation functions,
since distance between galaxies or pre-galaxies should expand with the
expansion of the universe. Can one find a discontinuity in the index of
correlation functions?

\subsection{Correlation functions in distance r and angular variable}

The correlation function is represented by a single power law\cite{peebles} in
the distance variable,%
\begin{equation}
\xi(r)=(\frac{r_{0}}{r})^{1.77\pm0.04}.
\end{equation}
However, the same book reported\cite{peebles2} a power law split in angular
correlation functions. These two sets of data may indicate that the distance
correlation functions may not reach to far distance, while the angular
correlation functions may have a far reach. Thus one may look for more recent
data from the galaxy redshift survey.

\subsection{ESO Slice Project (ESP) Galaxy Redshift Survey}

Recent data from the galaxy redshift survey\cite{redshsurvey} contains
information on correlation functions at further distance than that of the
distance correlation functions. They obtained a discontinuity in the power law
index in the red shift variable $hs$. This discontinuity at $hs=3(Mpc)$ may be
viewed as a split between radiation-dominated and matter-dominated expansion
rates. Obviously, the smaller vaue of $hs$ corresponds to radiation-dominated,
and the larger value to matter-dominated expansion. In other words, this is
the phenomenon in cosmology that corresponds to the knee energy in the cosmic
ray energy spectrum.

\section{Relation between AGN and galaxies}

In the old picture of black holes, AGN is the outcome of collapsed or collided
galaxies. Once the AGN stage is attained, that of a massive black hole, there
is no way to go back to an ordinary galaxy, since going to a smaller black
hole is prohibited. Then, why are most AGN observed at far distant locations?
In the new picture of black holes, however, this is not required. By collision
or collapse, a massive black hole can go to a smaller black hole or even to an
ordinary galaxy, if enough objects are emitted and low energy components are
accumulated at a nearby location. At least, that is an added scenario to play
with. It is quite conceivable that an object can start as a massive black
hole, i.e. an AGN object, then, by a collision or collapse of accumulated
matter in the neighborhood yield an ordinary galaxy. This possibility provides
flexibility in constructing added scenarios for galaxy formation in cosmology.

\section{Implication of the proposed model}

One of the most significant advancements in cosmic ray observations has been
the discovery of correlations between very high energy cosmic rays and AGN by
the Pierre Auger Observatory\cite{auger}. This may resolve the source of high
energy cosmic rays and clarify the mechanism of cosmic ray acceleration.
However, theoretical developments in that direction are yet to appear. The
model described in this article is to respond to such a requirement. A summary
is given in the following.

I). The author proposed a model in a series of papers in 1985, as quoted in
\cite{cr1}-\cite{cr9}. In other words, these articles have predicted the data
from the Pierre Auger Observatory.

II). It predicts a gravitational acceleration of cosmic rays from AGN, as well
as that of neutral particles, such as gamma rays, neutrinos and dark matter particles.

III). The knee energy is caused by a difference in the rates of
radiation-dominated and matter-dominated expansion. This requires the
existence of a particle at the knee energy scale, 3 PeV. That can be tested by
various kinds of cosmic ray detectors through the discovery of new particles
in the near future, which would be reminiscent of the golden age of cosmic ray
studies in 1950, when all new particles were discovered in cosmic ray events
before the accelerator age..

IV). The energy spectra of all other neutral particles should show the knee
energy phenomema at 3 PeV. This is an important prediction to be tested in the
near future.

V). Using the GLMR thoery, the mass of a dark matter particle has been
calculated from the knee energy mass scale of 3 PeV to be 8.1 TeV. This
matches the analysis of the HESS data which gives 7.6 $\pm$ 0.1 TeV.

VI). The explosion of supernovae can be explained by reasoning similar to that
of AGN explosion after gravitational collapse. This gives a unified reasoning
for the emission of cosmic rays from AGN and supernovae.

In summary, the proposed model explains the data from the Pierre Auger
Observatory successfully and many other predictions for observations will be
tested in the near future.

\bigskip

\begin{acknowledgments}
\bigskip The author would like to thank Lawrence W. Jones and Jean Krisch for
useful discussion and David N. Williams for reading the manuscript.
\end{acknowledgments}

\bigskip

\begin{figure}[h]
 \vspace*{-0.2cm}
 \begin{center}
 \includegraphics[width=0.6\textwidth]{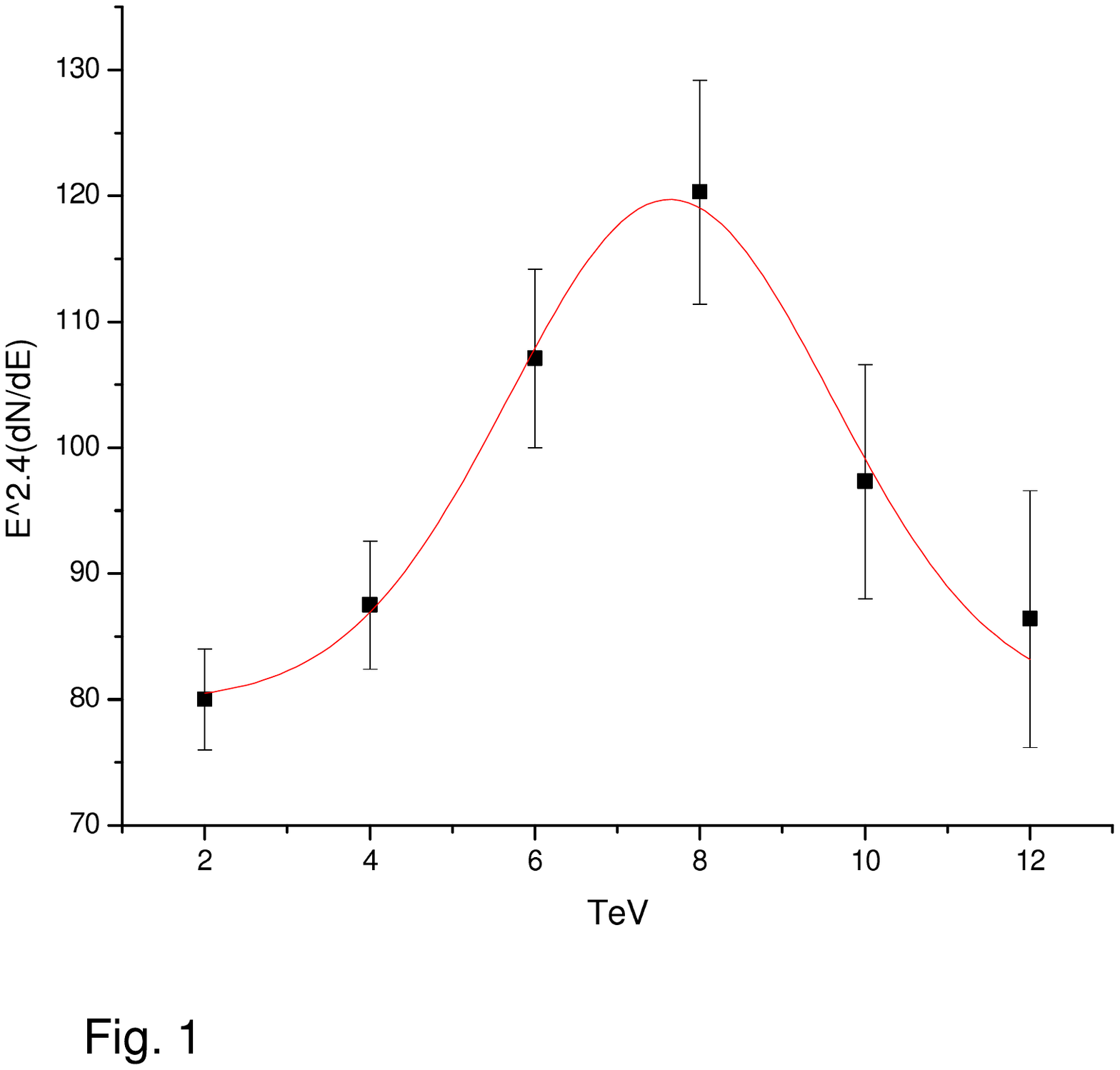}
 \caption{Sum of gamma ray energy spectra of 8 unidentified
sources\cite{hess8}. The y axis is $E^{2.4}(dN/dE)$ in units of $10^{-12}%
(TeV)^{0.4}(erg$ $cm^{-2}$ $s^{-1})$.}
 \label{fig1}
 \end{center}
 \end{figure}

\end{document}